# IA PARA EL MANTENIMIENTO PREDICTIVO EN CANTERAS: MODELADO

*F. Marcos-Macías[1]\*, R. Tamaki-Moreno[1], M. Cámara[1,2], V. Yagüe-Jiménez[1], J.L. Blanco-Murillo[1,2]*

[1] ETSI Telecomunicación, Universidad Politécnica de Madrid. Avda. Complutense 30, 28040 Madrid
[2] Information Processing and Telecommunication Center, UPM. Avda. Complutense 30, 28040 Madrid

**RESUMEN**

La dependencia de materias primas, especialmente en el sector minero, es una pieza clave en la economía actual. Los agregados son vitales, siendo la segunda materia prima más usada después del agua. Transformar digitalmente este sector es clave para optimizar las operaciones. Sin embargo, la supervisión y mantenimiento (predictivo y correctivo) son desafíos poco explorados en este sector, debido a las particularidades del sector, la maquinaria y las condiciones ambientes. Todo ello, a pesar de los éxitos alcanzados en la monitorización con sensores acústicos y de contacto.

Presentamos un esquema de aprendizaje no supervisado que entrena un modelo de autocodificador variacional sobre un conjunto de registros de sonido. Se trata del primer conjunto de datos de estas características recabado durante las operaciones de las plantas de procesamiento, y que contiene información de distintos puntos de la línea de procesado.

Los resultados evidencian la capacidad del modelo para reconstruir y representar en el espacio latente los sonidos registrados, las diferencias en las condiciones de operación y entre los distintos equipos. En el futuro, esto debe facilitar la clasificación de los sonidos, así como la detección de anomalías y patrones de degradación en el funcionamiento de la maquinaria.

**ABSTRACT**

Dependence on raw materials, especially in the mining sector, is a key part of today's economy. Aggregates are vital, being the second most used raw material after water. Digitally transforming this sector is key to optimizing operations. However, supervision and maintenance (predictive and corrective) are challenges little explored in this sector, due to the particularities of the sector, machinery and environmental conditions. All this, despite the successes achieved in other scenarios in monitoring with acoustic and contact sensors.

We present an unsupervised learning scheme that trains a variational autoencoder model on a set of sound records. This is the first such dataset collected during processing plant operations, containing information from different points of the processing line.

Our results demonstrate the model's ability to reconstruct and represent in latent space the recorded sounds, the differences in operating conditions and between different equipment. In the future, this should facilitate the classification of sounds, as well as the detection of anomalies and degradation patterns in the operation of the machinery.

***Palabras Clave—*** Artificial Intelligence, Predictive Maintenance, Noise Characterization, Self-supervised Learning, Variational Autoencoder.

## 1. INTRODUCCIÓN

Las materias primas desempeñan un papel fundamental en la economía y la industria. Actualmente, distintos países se enfrentan a un desafío constante por asegurar el suministro [1] y la sostenibilidad en la obtención, procesamiento y empleo de estas materias [2]. El objetivo es asegurar la competitividad dentro del mercado global, y promover el desarrollo económico y tecnológico que le permita asegurar el autoabastecimiento, así como una evolución continua y controlada de los costes de producción y los precios de venta [3], así como el reciclaje y la reutilización de los espacios [4].

La Unión Europea ha adoptado una estrategia integral que abarca desde la extracción y producción de materias primas hasta su reciclaje y reutilización [5]. Con ello, se busca minimizar los impactos ambientales y sociales negativos asociados con su producción y uso [6]. Ello permite optimizar el acceso a los recursos, reduciendo la dependencia de terceros países y fomentando la investigación y la innovación en tecnologías que permitan un uso más eficiente y la transición hacia una economía circular [7] [8].

---

\* ***Autor de contacto***: fernando.marcos.macias@alumnos.upm.es

Estos esfuerzos se alinean con los de la industria minera para dar respuesta al reto pendiente de la digitalización. El trabajo es particularmente intenso en el caso de los áridos, que constituyen la segunda materia prima más empleada después del agua, y son un componente básico en la construcción de edificios, carreteras e infraestructuras [9]. En este sector, el número de negocios familiares es alto, el precio de venta por unidad es bajo y el número de trabajadores en planta es cada vez menor [10], muchas se encuentras en localizaciones con conectividad limitada [11], cubren extensiones de terreno amplias y sufren de condiciones ambientales extremas: calor y frio, altos niveles de polvo, humedad, etc. [12].

La optimización en la gestión optimizada de los recursos, la monitorización del proceso productivo y la seguridad, son aspectos críticos para alcanzar un resultado favorable [13] [14]. El desarrollo y la proliferación de los sistemas de sensado autónomo, la llamada Internet de la Cosas (IoT, *Internet of Things*) y la inteligencia artificial (IA), están propiciando continuos cambios y avances. A pesar de ello, las características particulares del sector han hecho que este proceso esté siendo particularmente lento. Todavía son necesarias acciones conjuntas y desde las Administraciones, para fomentar el desarrollo industrial del sector y asegurar su continua evolución [5]. Proyectos como DigiEcoQuarry [15] buscan acelerar este proceso de transformación digital [11], manteniendo un enfoque transversal sobre todo el proceso de producción, y prestando especial atención a las necesidades de la industria y de todos los agentes involucrados.

Entre los varios puntos de trabajo en las operaciones de las plantas de procesado, los procedimientos orientados a la monitorización de la maquinaria [16] y el mantenimiento predictivo [17] [18] son algunos de los más relevantes y a los que se ha asignado un mayor potencial de impacto. Los métodos tradicionales para evaluar el estado y el funcionamiento presente y futuro de los sistemas industriales involucran el uso de técnicas de monitorización de la vibración, el rendimiento, o el sonido, entre otros. Estos enfoques requieren instalar sensores tanto de contacto, como acelerómetros, sondas de proximidad, transductores de presión y transductores de temperatura [19], como aéreos, como pueden ser los micrófonos o cámaras [20]. Estos segundos ofrecen grandes ventajas. Además de que no requieren contacto físico, su instalación se puede realizar de forma rápida y económica, evitando restricciones habituales en cuanto al espacio y el coste de sensores más caros. Todo ello es especialmente claro en el caso de los sensores acústicos, donde los componentes son fácilmente sustituibles y sus costes son bajos.

Las señales acústicas recopiladas contienen información relevante sobre la salud operativa de la máquina; pero también son sensibles al ruido de fondo y a los cambios en las condiciones de funcionamiento de la máquina. Esto ha sido ampliamente estudiado [19] en sectores como la automoción [21], la aviación o la hidráulica [22], alcanzando un buen número de casos de éxito. Sin embargo, ni la literatura científica ni las distribuidoras comerciales han registrado trabajos, estudios o productos similares en canteras. Tampoco orientados a la aplicación en la planta de procesado, ni a la aplicación de la IA y su integración.

Durante décadas el estudio de las señales acústicas en las canteras se ha centrado en el impacto acústico de estas instalaciones en el entorno o en sus operarios [23]. Esto resulta especialmente relevante allí donde se desarrollan voladuras, pero es igualmente crítico en cuanto a la maquinaria y los equipos.

En este sentido los trabajos científicos desarrollados con los operarios han evidenciado el valor de la información sonora en la gestión y operación de las canteras. Los propios operadores son los primeros en subrayar su valor para identificar y localizarlos problemas y fallos dentro de las instalaciones, así como reconocen la imposibilidad de mantener una monitorización suficiente de las plantas sobre esta información sin medios automáticos.

El objetivo de este trabajo es paliar esta carencia, desarrollando una primera aproximación a la tarea. La forma de hacerlo requiere empezar por recabar nuevos registros en planta y diseñar un esquema de modelado y caracterización apropiado. Esto debe permitirnos describir los sonidos registrados, representarlos convenientemente, y empezar a clasificarlos. El trabajo presenta los resultados de una primera aproximación al problema.

El trabajo se ha estructurado en siete secciones. La sección 2 repasa el estado del arte sobre la detección de anomalías mediante el procesado de registros acústicos, así como las técnicas de Inteligencia Artificial empleadas en estas tareas de representación y modelado. En la sección 3 se presenta el esquema beta-VAE propuesto para la caracterización y modelado de los registros, empleando representaciones del sonido que permiten reconstruir la fase. En la sección 4 se presentan los materiales empleados, en particular las bases de datos recabadas. En la sección 5 se describen los experimentos desarrollados, cuyos resultados se incluyen en la sección 6. Por último, la sección 7 incluye las conclusiones del trabajo y una lista de trabajos futuros a trabajar.

## 2. ESTADO DEL ARTE

La detección de anomalías acústicas con vistas a la detección de defectos en el funcionamiento y el mantenimiento predictivo de maquinaria es una tarea bien conocida y trabajada durante décadas. El estado del arte recoge extensos trabajos de aplicación [21] [22] [24]. Además, existen amplias revisiones acerca del estado del arte sobre las técnicas subyacentes para la detección de estas anomalías, que en buena medida podemos agrupar en dos bloques. Aquellas orientadas a la ingeniería de características (en inglés, *Feature Engineering*) y aquellas más recientes guiadas por los avances del aprendizaje automático, el aprendizaje profundo y la inteligencia artificial [26].

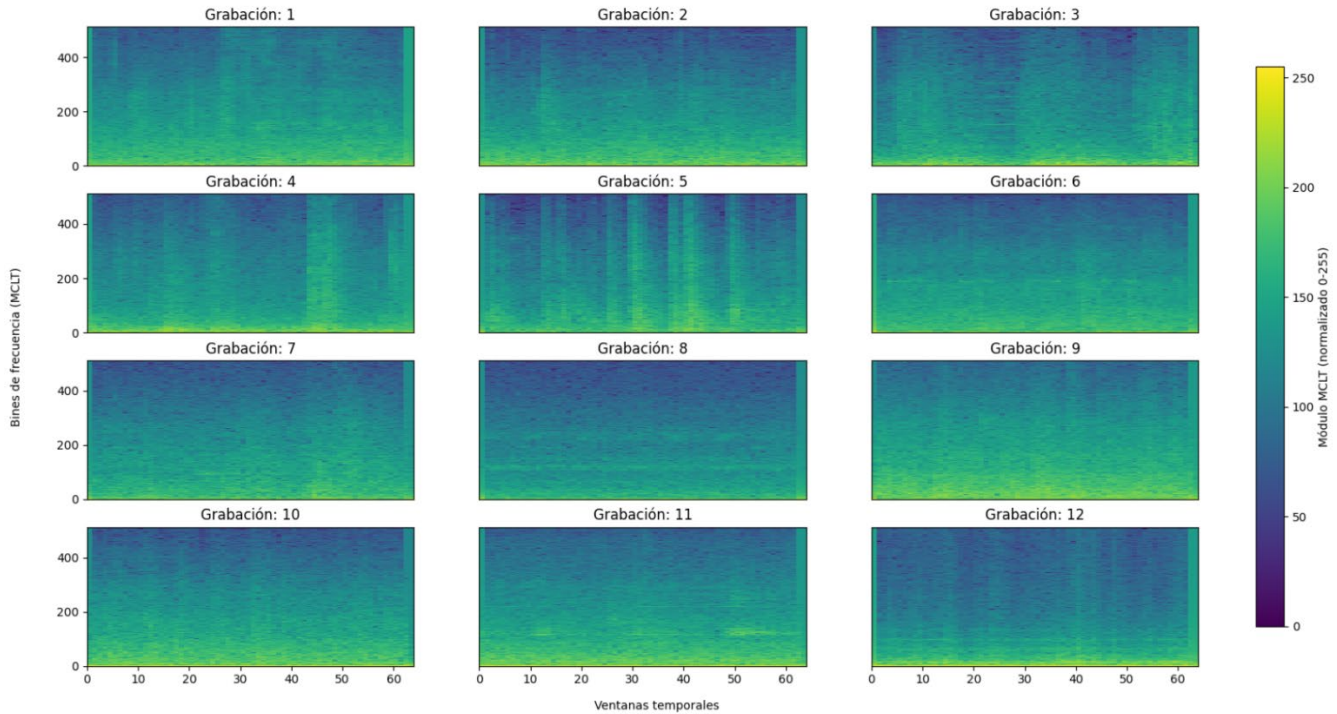

**Figura 1.** Representaciones MCLT de doce de las grabaciones registradas. En todas es posible observar el carácter ruidoso de los sonidos registrados, si bien es posible identificar ventanas asociables con golpes (gran ancho de banda, bandas verticales) y otras donde se observan componentes periódicas a lo largo del tiempo (bandas horizontales).

El paso de unas a otras es crítico, dado que las primeras se centran en desarrollar pruebas controladas para la sección de los parámetros de representación más adecuados. Típicamente, estadísticos de orden por ventanas o segmentos. En cambio, las segundas emplean representaciones convencionales (espectros, coeficientes MEL o BARK, etc.), y modelos avanzados para obtener nuevas formas de cuyas propiedades resulten más ajustadas a la tarea a desarrollar.

Al comparar ambas estrategias comprobamos que la primera presenta grandes ventajas en cuanto al número de parámetros, la posibilidad de trabajar con un número de datos relativamente menor, y la robustez frente al ruido. Por el contrario, adolece importantes limitaciones en cuanto a su capacidad de representación, dificultades para encontrar relaciones complejas y una dependencia excesiva de las decisiones tomadas por los expertos. En el caso de las técnicas no supervisadas, es posible alcanzar un buen resultado y una buena capacidad de representación, si bien el número de datos necesarios para ello es más elevado.

Los autocodificadores convencionales están limitados en cuanto a su capacidad para representar datos no observados, o en su defecto, alejados en un cierto sentido de las muestras empleadas en el entrenamiento. Esto restringe su uso en aplicaciones de mundo abierto, y sobre todo cuando el conjunto de datos de entrenamiento es muy limitado.

En este trabajo nos centramos en los autocodificadores variacionales partiendo de representaciones espectrales de registros segmentados uniformemente y sin etiquetas. Como hemos comentado, el empleo de autocodificadores para la detección de anomalías en procesos industriales no es nuevo [20]. Sin embargo, nunca se han empleado en su versión variacional para el modelado de señales de audio registradas en ambientes tan ruidosos, ni tampoco sobre registros estereofónicos. Todo esto resulta de especial interés cuando se emplean montajes en los que los micrófonos podría permitirnos una reconstrucción espacial mínima que permita localizar las fuentes sonoras [27], y cuando se busca tratar anomalías en registros cuya naturaleza es de por si ruidosa.

## 3. EL MODELADO DEL SONIDO

En esta sección presentamos el modelo diseñado y entrenado para recoger las características de los registros sonoros recabados en la cantera. Todo ello con vistas a la detección de patrones anómalos en su comportamiento o que en su condición pudieran dar lugar a variaciones en el sonido que puedan servir como predictores del estado de la operación o de los equipos industriales.

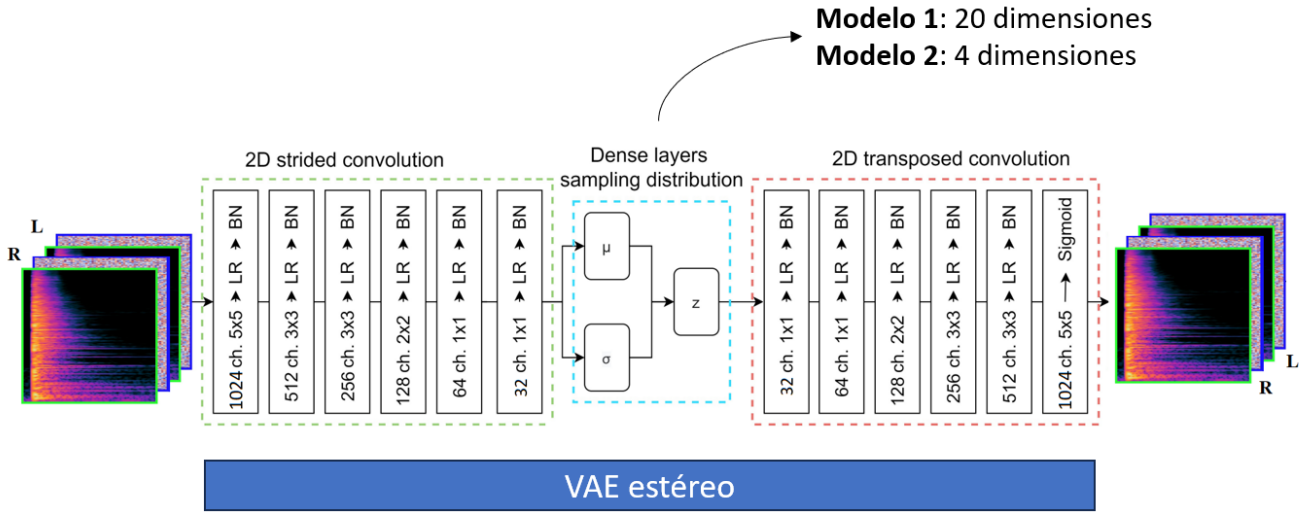

**Figura 2.** Esquema general del beta-VAE entrenado. El esquema incluye los dos canales estéreo (Right, Left) y las dos representaciones de módulo y fase para cada uno de ellos. Se han considerado hasta tres dimensiones para el espacio latente: modelo 1 (20 dimensiones), modelo 2 (4 dimensiones), modelo 3 (50 dimensiones).

El VAE es una arquitectura de aprendizaje profundo que codifica sus entradas en puntos de un nuevo espacio de variables latentes, y las decodifica de nuevo en señales de salida. En audio estas entradas y salidas suelen ser espectrogramas o representaciones similares basadas en el análisis frecuencial. Por el contrario, los vectores codificados del espacio latente, carecen de esta naturaleza y presentan una dimensionalidad mucho menor. El VAE se construye en una única fase siguiendo los mismos pasos que los demás autocodificadores, mientras que la codificación y la decodificación emplean la misma estructura de parámetros.

El entrenamiento de la red se centra en evaluar los valores de las media y varianzas del modelo normal estándar impuesto al espacio latente en su distribución. En este punto, el esquema variacional puede incorporar términos adicionales en el error que permitan controlar su aprendizaje. El caso más sencillo y popular es el del beta-VAE, que toma su nombre del parámetro β que controla el aprendizaje ponderando la función de error del VAE con la divergencia Kullback-Leibler en la distribución asignada al espacio latente. Los detalles de las distintas implementaciones están ampliamente descritos en la literatura, siendo crítico conocer la naturaleza probabilística del entrenamiento y la naturaleza multidimensional de la representación latente [28].

Para el modelado de registros con varios canales cabe introducir importantes cambios en la estructura habitual de un VAE. Por ejemplo, podemos considerar por separado cada uno de los canales, evaluando varios VAEs monoaurales, en paralelo pero sin interacción, o en conjunto, permitiendo la interacción entre los distintos canales. Esencialmente, la estructura de dos VAEs monoaurales duplica el número de coeficientes, mientras que un esquema estereofónico debe ser capaz de trabajar sobre una única representación, en un único espacio latente. En todo caso, el esquema estéreo emplea información mínima de etiquetas que permiten mantener separados los dos canales (derecho e izquierdo).

La Fig. 2 ilustra el modelo de beta-VAE propuesto, incluyendo los dos canales de audio (derecho e izquierdo) y dos representaciones para cada uno, correspondientes al contenido de amplitud y fase. La importancia de esta separación ha quedado demostrada en trabajos anteriores, con vistas a la reducción del error de reconstrucción y a un mejor tratamiento del contenido sonoro de los registros [29].

El entrenamiento del VAE minimiza el error de reconstrucción de las salidas respecto de las entradas, lo que nos permite centrarnos en primera instancia en el modelado de los registros, sin necesidad de abordar otros aspectos tales como la posterior capacidad de detección de efectos espurios. De esta forma, asumimos que estas posibles anomalías tendrán una representación diferente al resto en el espacio latente. Esto nos permitirá posteriormente identificarlas y detectarlas, sencillamente por el reflejo en el espacio latente de las diferencias en el contenido acústico. Como consecuencia, podemos centrarnos en primera instancia en el entrenamiento de los modelos, debiendo evaluar en último término no sólo las reconstrucciones (la capacidad de modelado) o la capacidad generativa, sino también las representaciones asociadas a posibles anomalías.

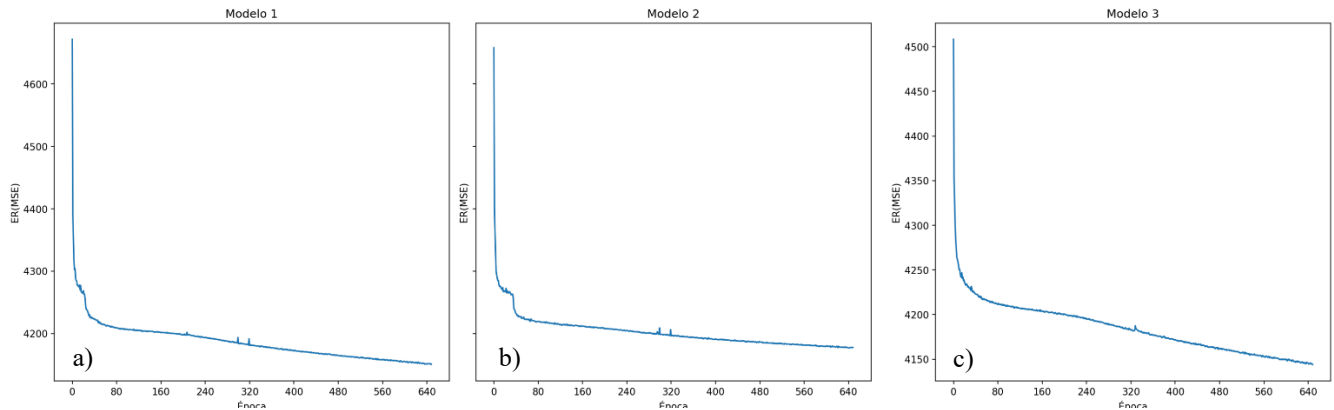

**Figura 3.** Resultados del entrenamiento de los tres modelos a lo largo de las distintas épocas. En todos los casos se observa un rápido descenso en el error cuadrático medio al inicio y un decaimiento sostenido al final.

Algunos trabajos previos han adaptado la métrica de error a las métricas perceptuales que imitan el comportamiento del oído humano. En nuestro caso trabajamos con el error cuadrático medio, como se recoge habitualmente en la literatura, evitando sobrecostes computacionales.

Adicionalmente a todo lo anterior, la literatura recoge alternativas de interés cuando se quiere abordar un posterior proceso de clasificación o de detección (i.e., clasificación en dos clases). El empleo de estructuras adversarias después de entrenar el VAE y sólo para corregir el decodificador parece haber permitido alcanzar importantes mejoras en la calidad de las reconstrucciones de secuencias continuas, sin lastrar el entrenamiento inicial del VAE [30]. Esta posibilidad invita a pensar en estructuras alternativas a la propuesta, y en futuras mejoras una vez que se logre modelar el contenido acústico de los registros de manera no supervisada.

## 4. MATERIALES

Para este trabajo hemos recabado un conjunto de registros que contiene más de 4 horas de grabaciones recogidas en distintos puntos de tres canteras diferentes. En cada una se tomaron registros de más de 20 minutos cubriendo áreas donde operaban cuatro componentes principales: cintas, molinos, machacadoras y cribas. En este sentido, es importante destacar las características ambientales de las canteras de áridos pueden ser muy variadas, pero en general destacan los altos niveles de polvo (particularmente en explotaciones de extracción fuera del agua), los niveles de humedad (en graveras), la variabilidad del viento en su dirección e intensidad, amplia excursión a lo largo de escalas de temperatura en función del clima y la estación, etc. Todo esto hace necesario repetir las medidas en nuevos entornos, y tomar precauciones para evitar dañar los equipos.

La captura se realizó con una grabadora ZOOM H8 de ocho pistas independientes y con autonomía de más de diez horas. Ésta tenía acoplada una cápsula X/Y XYH6 que emplea dos micrófonos unidireccionales de alta calidad colocados en ángulo. Esta configuración es óptima para cubrir un área amplia y aún así capturar una potente imagen central. En este caso, se evitó proteger los micrófonos con ningún elemento que pudiera afectar la calidad de las señales registradas o modificar su contenido frecuencial. Trabajos posteriores deberán evaluar la necesidad de incluir estos elementos, particularmente en instalaciones permanentes.

La grabadora permitió recoger registros completos de más de media hora, permitiendo además la comprobación *in situ* de la calidad de los audios. Posteriormente, todos los registros fueron fragmentados a una duración de segmento igual, conservando únicamente las etiquetas de los distintos ficheros para posteriores comprobaciones. La información relativa a los elementos de la planta de producción en cuyo entorno se situó la grabadora no se ha empleado en el entrenamiento, si bien será tenida en cuenta en la evaluación del modelo entrenado (ver Sección 6).

La representación elegida para los fragmentos de audio recortados corresponde a los coeficientes de la Transformada Modificada del Coseno, MCLT. El motivo detrás de esta elección viene justificado por trabajos previos que ilustran la importancia de cuidar la información de fase, aun cuando el contenido frecuencial de los registros sea ruidoso [29].

La Fig. 1 muestra a modo de ejemplo las representaciones MCLT de doce segmentos extraídos de registros distintos de la base de datos. Cada uno de los segmentos contiene 31,994 muestras, mientras que todas las grabaciones fueron registradas a 48 kHz de frecuencia de muestreo y con 24 bits de profundidad en cada muestra. Esta misma representación se empleó en todos los ficheros registrados.

## 5. EXPERIMENTOS

Hemos desarrollado cuatro experimentos para validar el comportamiento del modelo entrenado, y considerado tres variaciones sobre el esquema de VAE estéreo. Los tres modelos emplean el mismo esquema, variando el número de dimensiones del espacio latente. Hemos considerado tres valores, 20, 4 y 50 dimensiones, en base a nuestra experiencia previa y ensayos preliminares. Un estudio exhaustivo de este parámetro podría resultar en una mejora poco significativa de los resultados reportados en este trabajo, pero en ningún caso debe ofrecer nueva luz sobre el sistema desarrollado.

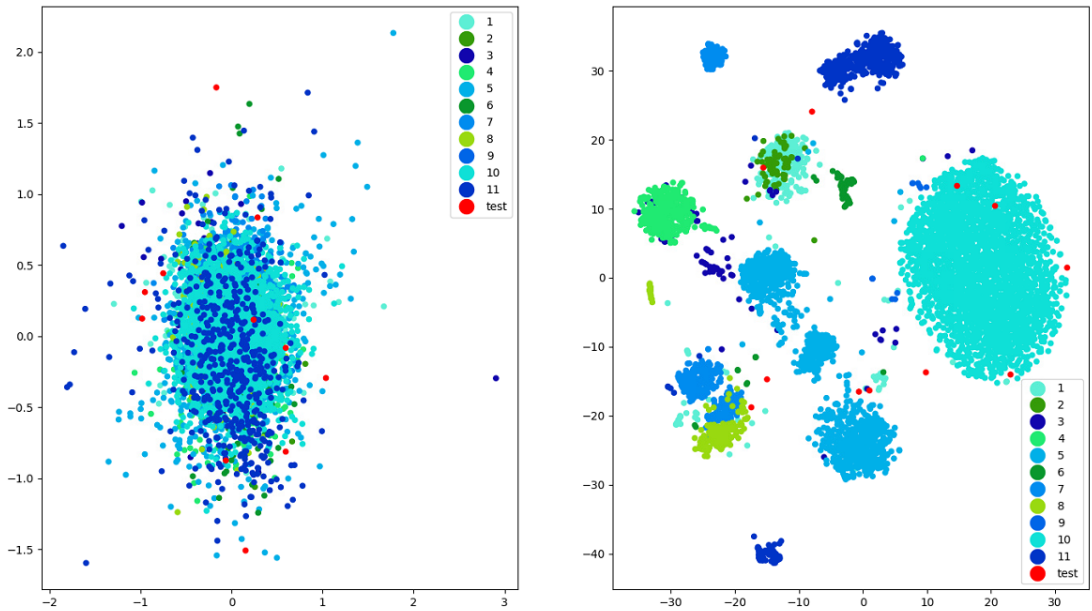

**Figura 4.** Representaciones del espacio latente obtenida para el modelo 1 (20 dimensiones): a) PCA (izquierda), b) t-SNE (derecha). Los colores codifican los distintos registros de audio, se representa un punto por cada fragmento recortado. Los puntos en rojo representan datos anómalos, importados para evaluar el comportamiento de los modelos entrenados.

La Fig. 3 ofrece una vista general sobre el proceso de entrenamiento de los tres modelos en base al error cuadrático medio en la reconstrucción. Los tres modelos evidenciaron un rápido proceso de descenso del error al principio, y presentan una zona de decaimiento más lento y sostenido al final del entrenamiento que no llega a culminar. Esto nos hace pensar que los modelos obtenidos pueden servir como referencia en investigaciones posteriores centradas en la mejora del proceso de entrenamiento. En todo caso, resulta interesante comprobar que la tendencia que se observa al final es más abrupta a medida que aumenta el número de dimensiones del espacio latente, pero en ningún caso de una forma determinante.

En vista de los modelos obtenidos, a continuación, describimos los cuatro experimentos desarrollados a fin de evaluar la calidad del modelado realizado. Cada uno de ellos aborda un aspecto diferente del modelo resultante, empezando por el error de reconstrucción a la salida del VAE hasta la estructura del espacio latente y un estudio preliminar sobre la capacidad del esquema propuesto para representar anomalías.

### 5.1. Evaluación del error de reconstrucción

Esta evaluación involucra únicamente el error cuadrático medio que se obtiene en la reconstrucción de las entradas a la salida del VAE. Esta métrica coincide con la que se emplea en el proceso de entrenamiento, lo que nos permite comprobar el comportamiento de los modelos finalmente seleccionados.

En este caso, los modelos seleccionados no son los que se obtienen al final del entrenamiento, sino los que se tienen al inicio de la cuesta de decaimiento sostenido del error. De esta forma podemos comparar los tres modelos del esquema propuesto en condiciones similares.

### 5.2. Evaluación de la calidad de reconstrucción

Además de considerar el error cuadrático medio, resulta muy relevante evaluar otras métricas de error que puedan incorporar aspectos perceptuales. Tal y como hemos explicado, en la práctica, sabemos que los operarios de las canteras se sirven de la información acústica para monitorizar las operaciones, detectar problemas e incluso para verificar que han sido correctamente resueltos.

Por esta causa, hemos comparado los resultados del error cuadrático medio con los valores MOS de PEAQ y ViSQOL tanto en la media como en su dispersión. Estos valores nos dan una idea de la capacidad de representación del modelo entrenado y del error esperable.

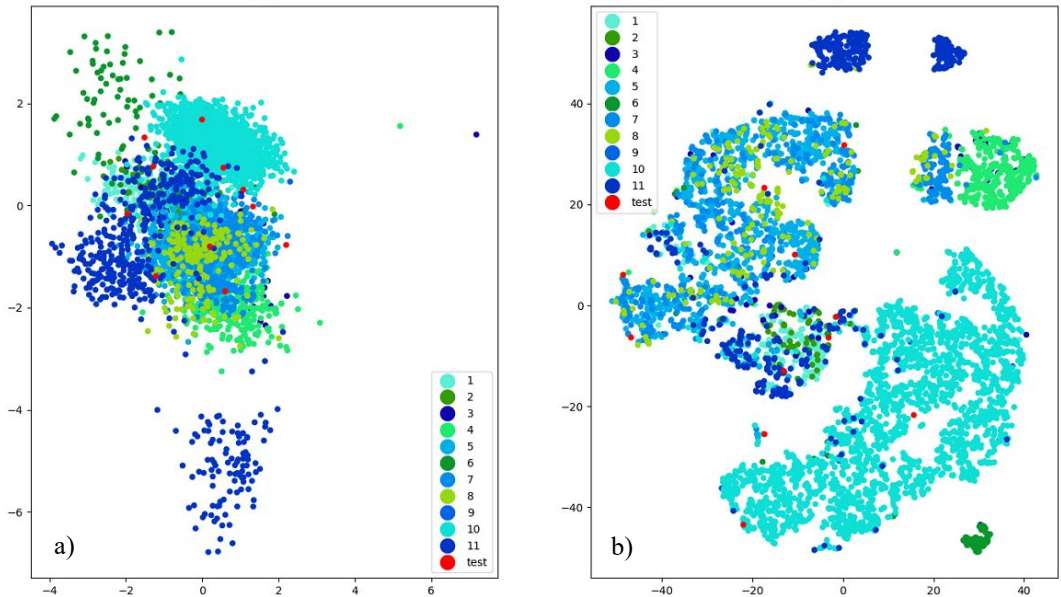

**Figura 5.** Representaciones del espacio latente obtenida para el modelo 2 (4 dimensiones): a) PCA (izquierda), b) t-SNE (derecha). Los colores se codifican de la misma forma que los de la Fig. 4.

Adicionalmente a las pruebas realizadas sobre métricas objetivas que aproximan a las subjetivas, los audios generados se encuentran disponibles en abierto para su consulta[2]. Es muy recomendable dedicar unos minutos a escuchar tanto los registros originales grabados como sus reconstrucciones a fin de valorar la calidad del modelado y comprender la dimensión del reto que abordamos.

### 5.3. Evaluación del espacio latente: PCA & t-SNE

Dadas las características de los audios que nos ocupan (fundamentalmente ruidosos) es importante considerar no sólo la calidad de la reconstrucción que ofrece nuestro modelo no supervisado, sino también las características del espacio latente. Es bien sabido que las propiedades de los datos procesados por el autocodificador van a reflejarse o bien en su estructura (pesos de la red) o en las propiedades de su representación codificada (espacio latente). Las segundas son particularmente interesantes en nuestro caso, puesto que se centran en las variaciones más representativas de los sonidos, y por tanto, son fuentes de información relevante.

En este caso, dada la dimensionalidad del espacio latente entrenado (desde 4 dimensiones hasta 50), es necesario emplear algoritmos de reducción de la dimensionalidad como PCA (análisis de componentes principales) o t-SNE (*t-distributed Stochastic Neighbor Embedding*). Ambos son herramientas de uso habitual para la visualización del espacio latente. El primero se centra en las propiedades generales del espacio e identifica dimensiones ortogonales donde se concentra la varianza. El segundo se centra en sus propiedades locales de los datos proyectados en el espacio latente. En particular, trata de preservar las distancias entre los vecinos en el espacio latente multidimensional, tras la proyección en el espacio de menor dimensión y mayor inteligibilidad. Tanto para PCA como para t-SNE empleamos una reducción a dimensión dos lo que debe facilitar la visualización de los datos.

### 5.4. Estudio preliminar de anomalías

Finalmente, a modo de ejemplo, hemos analizado la presencia de anomalías en los conjuntos de datos. Para ellos hemos asumido que los conjuntos recabados, homogéneos o no, representan una situación de operación normal en la cantera, y que por tanto, no contenían anomalías severas. Bajo esta hipótesis, hemos añadido fragmentos de un audio que registra una grabación de una pieza de maquinaria dañada. Con vistas al sistema de mantenimiento predictivo y a la detección de anomalías, cabe esperar que los fragmentos de este registro tengan un comportamiento diferencial comparado con los recabados en la cantera.

Los fragmentos evaluados, cumplen con los requisitos de las restantes grabaciones (duración de los segmentos, frecuencia de muestreo, precisión en la codificación, etc.), pero no fueron registrados con el mismo dispositivo ni en el mismo ambiente acústico. Por tanto, cabe considerar que se trata de un elemento fuera del dominio. Sin embargo, se ha podido verificar manualmente que su contenido se ajusta al del ruido de maquinaria con defectos en su funcionamiento, asumiendo que éste debe poder tener un patrón característico en su representación latente.

---

[2] https:// fermarcosmac.github.io/vae-estereo-canteras/

Este patrón característico debe ser identificable en el espacio latente de los modelos. Dada la elevada dimensionalidad de este espacio volvemos a emplear las visualizaciones de la subsección anterior (PCA y t-SNE), incorporando ahora los datos correspondientes a las proyecciones de los segmentos del nuevo audio en el espacio latente. En este caso, dada la naturaleza de los datos, podemos sencillamente considerar ambos experimentos de forma conjunta, y representar sus resultados agregados, incluyendo los registros de la base de datos y los fragmentos anómalos.

# 6. RESULTADOS

A continuación detallamos las figuras y tablas muestran los resultados de las distintas evaluaciones siguiendo el orden de la Sección 5. En primer lugar, la Tabla 1 resume los resultados obtenidos en la reconstrucción con los tres modelos y para las tres métricas consideradas: error cuadrático medio, ViSQOL y PEAQ. Los resultados obtenidos evidencian cómo la media y la varianza del error cuadrático medio están muy próximas, pero ambos valores son bajos. Esto supone en la práctica que el error de reconstrucción sea relativamente bajo, lo que puede ser consistente con la naturaleza ruidosa de los registros.

**Tabla 1.** Resumen de los resultados por los tres modelos entrenados sobre las métricas de calidad objetiva evaluadas: error cuadrático medio (MSE), ViSQOL y PEAQ.

| Modelo | Estadístico | MSE | ViSQOL | PEAQ |
|---|---|---|---|---|
| M1 | Media | **1.58 · 10⁻⁶** | **3.17** | -3.70 |
|  | Varianza | 1.55 · 10⁻⁶ | 0.34 | 0.44 |
| M2 | Media | 1.73 · 10⁻⁶ | 3.16 | -3.48 |
|  | Varianza | 1.51 · 10⁻⁶ | 0.36 | 0.63 |
| M3 | Media | 1.78 · 10⁻⁶ | 3.14 | **-3.75** |
|  | Varianza | 1.52 · 10⁻⁶ | 0.35 | 0.38 |

Por otro lado, los resultados obtenidos en las métricas ViSQOL y PEAQ difieren entre sí. En esta ocasión ViSQOL ofrece unos resultados bastante positivos, mientras que los de PEAQ son peores. Además, la varianza es sostenidamente menor para ViSQOL que para PEAQ. En este caso, es importante recalcar que la naturaleza ruidosa de los registros contrasta en cierta medida con el tipo de grabaciones que habitualmente evalúa ViSQOL, esto, unido a una evaluación subjetiva de los registros generados nos hace pensar que la calidad de la reconstrucción es todavía muy limitada, aún para el Modelo 3, con 50 dimensiones en su espacio latente.

En cuanto a la complejidad del modelo, sorprende que los valores obtenidos en las tres métricas sean análogos para un número de dimensiones tan dispar (de 4 a 50). Esto, ligado a las curvas de la Fig. 3 nos hace pensar en la necesidad de profundizar en las representaciones, abordando alternativas más centradas en la naturaleza ruidosa de los registros.

Una vez analizadas las métricas de calidad en reconstrucción abordamos la estructura del espacio latente de los modelos entrenados. En este caso hemos incluido los resultados del PCA y del t-SNE para los modelos 1 y 2 en las Figs. 4 y 5, respectivamente. Los resultados del modelo 3 (50 dimensiones) fueron totalmente análogos a los del modelo 1 (20 dimensiones), por lo que no se han incluido.

Las representaciones aportadas no solo muestran los datos sobre dos espacios de dimensionalidad reducida, sino que además representan en color los distintos registros que se están representando. Cabe recordar que en nuestro conjunto de datos cada registro corresponde a un punto de grabación, y potencialmente con una instalación diferente, lo que supone que estemos analizando audios asociados con distintos elementos de la planta de procesado, en distintos momentos e incluso en distintas localizaciones. En este caso hemos incluido no sólo fragmentos de los registros recabados en las canteras (Experimento 3), sino también fragmentos de un registro anómalo tomado de otro conjunto de datos (Experimento 4). Estos últimos han sido representado en rojo para facilitar su visualización.

A la izquierda de cada una de las figuras se observa el análisis de componentes principales reducido a dos dimensiones. En este caso, el modelo 2 (Fig. 5.a) parece representar dos clústeres diferenciados mientras que el modelo 1 (Fig. 4.a) sólo contendría uno. En este caso, la estructura del beta-VAE sugiere que la estructura del espacio latente debe ser normal, lo que estaría en consonancia con esta figura. El hecho de que los fragmentos de los distintos registros se entremezclen también sería consistente con esta estructura normal de los datos, en vista de que la proporción de fragmentos debe ser análoga para todos los registros.

En cuanto a los t-SNE, difieren considerablemente. El correspondiente al modelo 2, de menor dimensionalidad en el espacio latente, Fig. 5.b, entremezcla los distintos registros, aun cuando da lugar a formas particulares que evidencian la existencia de una estructura subyacente en el espacio latente. Por el contrario, el correspondiente al modelo 1, Fig. 4.a, logra separar en clústeres los distintos registros.

Finalmente, en cuanto a la anomalía, las representaciones obtenidas no separan claramente las proyectos de los fragmentos asociados, si bien los ubican en posiciones muy dispares dentro del espacio latente, lo que puede servir en un futuro para su detección.

## 7. CONCLUSIONES

En este trabajo hemos presentado un esquema de modelado de audio basado en un autocodificador variacional para el mantenimiento predictivo en canteras. El esquema tiene la estructura de un beta-VAE y fue capaz de representar los distintos sonidos recabados en las plantas de procesado de distintas canteras.

El entrenamiento desarrollado fue capaz de ofrecer un modelo capaz de regenerar los sonidos, así como también de generar nuevos sonidos a través del muestreo del espacio latente del VAE. El esquema evaluado emplea representaciones basadas en la MCLT, a fin de mejorar el tratamiento habitual que dan los esquemas basados en la magnitud de la transformada local de Fourier.

El error en la reconstrucción de los audios en base al modelo entrenado es bajo entorno a $10^{-6}$ en error absoluto y aceptable (MOS=3) cuando empleamos la métrica de evaluación de calidad ViSQOL. En cambio, PEAQ ofrece todavía resultados bajos, consistentes con lo observado tras las escucha de los audios reconstruidos.

Se trata del primer trabajo de estas características publicado, y el primero centrado específicamente en las líneas de procesamiento de canteras de áridos. El trabajo emplea un nuevo conjunto de datos recabado en distintas canteras, dentro de las actividades del proyecto H2020 DigiEcoQuarry, capturado en tres canteras de áridos en Italia (HOLCIM, Pioltello), Francia (VICAT GRANULATS, Toulouse), y España (HANSON, Valdilecha). Cada una de ellas presenta características operativas propias (gravera, de recuperación, y de explosión en superficie), pero incorporan elementos comunes en las líneas de procesado: tolvas, molinos, cintas, machacadoras y cribas, principalmente.

El esquema desarrollado se ha centrado en molinos, cintas, machacadoras y cribas, sin emplear ningún tipo de etiquetado en el entrenamiento. Este trabajo demuestra que un esquema de aprendizaje no supervisado como el descrito logra captar y registrar las características del ambiente sonoro, lo que ofrece importantes ventajas a la hora de desarrollar un esquema predictivo. Entre otras, evita la necesidad de etiquetar los datos a priori, permite comprimir los registros de audio hasta la dimensión asignada al espacio latente, y permite al esquema de inteligencia artificial operar de forma autónoma.

Trabajos futuros deberán analizar las variaciones en los datos cuando se emplean sistemas de grabación distintos, y la capacidad de estos esquemas para aprender los rasgos característicos del entorno. Todo ello aun cuando pueda tratarse de los mismos modelos de los equipos, o el ambiente sonoro pueda variar en función de la distribución espacial de las líneas de procesado en las canteras.

El análisis de las representaciones capturadas en estos modelos debe servirnos para desarrollar detectores especializados. Para esto será necesario evaluar la capacidad de los modelos para describir efectos o cambios que puedan representar anomalías en los sonidos. Estamos particularmente interesados en aquellos que, en última instancia, puedan ser asociados a desviaciones en el funcionamiento de los elementos de las líneas de producción, o que puedan ser buenos predictores de aquellos.

Este es el primer paso hacia el mantenimiento predictivo y automatizado en las canteras a través de registros sonoros. El sistema final deberá anticiparse a los fallos mecánicos y del funcionamiento, con una ventana temporal de predicción suficiente para ofrecer valor real a la industria. Todo ello deberá ser valorado a medida que se vayan completando los estudios necesarios, y se amplíe la base de información disponible.

## 8. AGRADECIMIENTOS

Este trabajo ha sido financiado conjuntamente por el Ministerio de Economía y Competitividad del Gobierno de España dentro del proyecto PID2021-128469OB-I00, el Programa de Investigación e Innovación de la Unión Europea Horizon 2020 dentro del "Grant Agreement No. 101003750".

Los autores agradecen expresamente la colaboración de las empresas operadoras de las canteras involucrados en este proyecto, así como a su personal técnico y humano por la colaboración prestada. En este caso a las canteras de HOLCIM-Piotello (Italia), por las pruebas preliminares realizadas allí en noviembre de 2022, HANSON-Valdilecha (España) y VICAT Granulats-Toulouse (Francia), por permitir el acceso a sus instalaciones y ayudar con las grabaciones de audio, en mayo y junio de 2023.

Los autores también quieren expresar su agradecimiento al Prof. José Eugenio Ortiz, por gestionar las grabaciones de audio en VICAT y liderar al equipo UPM-AI en DEQ, a los profesores María Josefa Herrero, Jose Ignacio Escavy e Iván Cabria, por su apoyo y participación en este proceso y en el desarrollo del servicio de Mantenimiento Predictivo. Finalmente, al científico Francisco Simón, del Consejo Superior de Investigadores Científicas, que facilitó el equipamiento para la primera captura de datos.

Finalmente, a la empresa Sigma Technologies y a su increíble equipo por organizar las visitas, coordinar y participar en el desarrollo de los servicios de IA DEQ: Pierre, Silvia, Javier, Cecilia, César.